\documentclass[11pt,a4paper]{article}
\usepackage[hyperref]{acl2020}
\usepackage{times}
\usepackage{latexsym}
\usepackage{graphicx}
\usepackage{amssymb}

\aclfinalcopy 

\begin{document}
\title{A Survey of Knowledge Graph Embedding and Their Applications}

\author{Shivani Choudhary, Tarun Luthra, Ashima Mittal, Rajat Singh \\
  Indian Institute of Technology Delhi \\
  Hauz Khas, Delhi-110016, India\\ \texttt{shivani@sire.iitd.ac.in},\\
  \texttt{\{mcs202475,anz208486,csz208507\}@cse.iitd.ac.in}}

\date{}

\maketitle
\begin{abstract}
Knowledge Graph embedding provides a versatile technique for representing knowledge.  These techniques can be used in a variety of applications such as completion of knowledge graph to predict missing information, recommender systems, question answering, query expansion, etc. The information embedded in Knowledge graph though being structured is challenging to consume in a real-world application. Knowledge graph embedding enables the real-world application to consume information to improve performance. Knowledge graph embedding is an active research area. Most of the embedding methods focus on structure-based information. Recent research has extended the boundary to include text-based information and image-based information in entity embedding. Efforts have been made to enhance the representation with context information. This paper introduces growth in the field of KG embedding from simple translation-based models to enrichment-based models. This paper includes the utility of the Knowledge graph in real-world applications.  
\end{abstract}



\section{Introduction}

Knowledge graph(KG) has received a lot of traction in the recent past and led to much research in this area. Most of the research is focused on the generation of Knowledge graphs and consumption of the information enshrined in the Knowledge Graph. Some of the earlier works to create KG are YAGO\citep{Suchanek2007}, Freebase\citep{Bollacker2008}, DBpedia\citep{Lehmann2012} and WikiData\citep{Vrandecic2014}. The evolution of the Knowledge Graph starts with the seminal paper from Berners-Lee \citep{Berners-Lee2001}. The knowledge graph has evolved in three phases. In the first phase, the Knowledge representation was brought to the level of Web standard. The core focus shifted to Data management, linked data, and its application in the second phase. In the third phase, the focus shifted on the real-world application \citep{Bonatti2018KnowledgeGN}. The real-world application ranges from Semantic parsing\citep{Berant2013, Heck2014}, recommender system\citep{Sun2020, Wang}, question answering\citep{ Saxena}, named entity disambiguation\citep{Lin}, information extraction \citep{Xiong, Xiong, Liu2018a, Dietz2019} etc. The knowledge graph is a representation of structured relational information in the form of Entities and relations between them. It is a multi-relational graph where nodes are entities and edges are relations. Entities are real-world objects or abstract information. The representation entities and relations between them are represented as triple. For e.g. \textit{(New Delhi, IsCapitalOf, India)} is an example of a triple. \textit{New Delhi} and \textit{Delhi} are entities and \textit{IsCapitalOf} is relation. Though the representation looks scientific but consuming these in the real-world application is not an easy task. The consumption of information enshrined in the knowledge graph will be very easy if it can be converted to numerical representation. \textit{Knowledge graph embedding} is a solution to incorporate the knowledge from the knowledge graph in a real-world application. \\
The motivation behind Knowledge graph embedding \citep{Bordes} is to preserve the structural information, i.e., the relation between entities, and represent it in some vector space. This makes it easier to manipulate the information. Most of the work in \textit{Knowledge graph embedding \textbf{(KGE)}} is focused on generating a continuous vector representation for entities and relations and apply a relationship reasoning on the embedding. The relationship reasoning is supposed to optimize some scoring function to learn the embedding. Researchers have used different approaches to learn the embedding Path-based learning\citep{Toutanova2016}, Entity-based learning, textual-based learning, etc. A lot of work was focused were translation model \citep{Bordes} and semantic-based model\citep{Bordes2014}. The representation of the triple results in a lot of information lost because it fails to take "textual information" into account. With the proposal of Graph attention network \citep{Velickovic2017a} the representation of the entities has become more contextualized. In recent years, the proposal of multi-modal graph has extended the spectrum to a new level. In multi-modal knowledge graph, Knowledge graph can have multi-modal information like image and text etc. \citep{Sun2020, Wei2019}. \\
Previous work on survey has focused on the KG Embedding \citep{Wang}, KG Embedding and application\citep{Ji2020b} , KG embedding with textual data \citep{Lu2020a}, KG embedding based on the deep-learning \citep{Wang2020a}. This work shall focus on the KG embedding from translation-based model, semantic-based model, embedding with enriched representation from textual data and multi-modal data and their application. In section 2, we shall provide the details of KGE; in section 3, we shall present the application area. In the summary section, we shall try to put emerging areas of research in KGE. 

\section{Knowledge Graph embedding}
\label{sec:length}

Knowledge Graph embedding is an approach to transform Knowledge Graphs (nodes, edges, and their feature vectors) into a low dimensional continuous vector space that preserves various graph structure, information, etc. These approaches are broadly classified into two groups: $\textit{translation models}$ and $\textit{semantic matching models}$.

\subsection{Translation Models}
The translation-based model uses distance-based measures to generate the similarity score for a pair of entities and their relationships. The translation-based model aims to find a vector representation of entities with relation to the translation of the entities. It maps entities to a low-dimensional vector space. 
\subsubsection{TransE \citep{Bordes2014}}
\begin{figure}
    \centering
    \includegraphics[width=.9\linewidth]{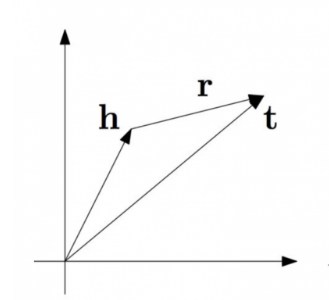}
    \caption{The intuition of TransE model. Image credit \cite{Wang}}
    \label{fig:my_label}
\end{figure}

The first model proposed was TransE. It is an energy based model. If a triple \textit{h, r, t} holds then the vector representation \textit{h} and \textit{r} should be as close as possible. It can be graphically represented as Figure-\ref{fig:my_label}. Mathematically, it can be stated like $\textit{h} + \textit{r} \approx \textit{t}$. The energy of a triple is  \textit{d(h + r,t)} for some similarity measure d. To learn the embedding, minimization of ranking based loss function over the training set.
\begin{equation}
    \mathcal{L= \sum_{\textit{h,r,t} \in S}  \sum_{\textit{$\hat{h}$,r,$\hat{t}$} \in \hat{S} } [\gamma+\textit{d(h+r),t}-\textit{d($\hat{h}$,r,$\hat{t}$)}]}
\end{equation}
\textit{d(\^{h},r,\^{t})} represents the set of the corrupt triples. This loss function will be optimized so that the valid triples are ranked above the corrupt triples. 
This model fails in case of the one to many relation and many to many relation. To overcome this deficit new model TransH is proposed.

\subsubsection{TransH \cite{Wang2014KnowledgeGE}}
\begin{figure}
    \centering
    \includegraphics[width=.9\linewidth]{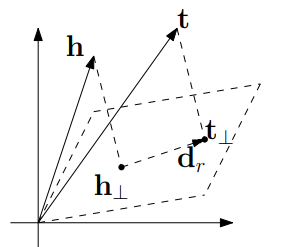}
    \caption{TransH model. Image taken from \cite{Wang}}
    \label{fig:my_label_2}
\end{figure}

It was proposed to address the limitations of TransE. This model enables an entity to
have distributed representations based on their involvement in the relation. The representation of \textit{$h_\bot$}, \textit{$t_\bot$} are projected in a relation specific hyperplane. The relation between then is \textit{$d_r$}. As in the Figure-\ref{fig:my_label_2}, the vectors \textit{h} and \textit{t} are projected in the relation hyper-plane. The loss function and intuition remain similar to TransE. 

\subsubsection{TransR \cite{10.5555/2886521.2886624}}
\begin{figure}
    \centering
    \includegraphics[width=.9\linewidth]{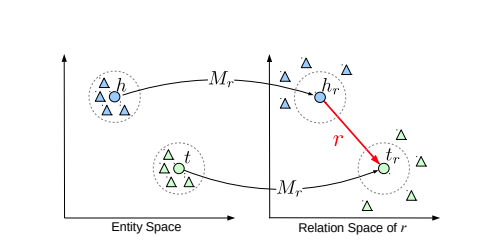}
    \caption{TransR model. Image taken from \cite{10.5555/2886521.2886624}}
    \label{fig:my_label_3}
\end{figure}

It proposes that an entity may have multiple attributes and various relations. Each relation may focus on different attributes of entities. TransR models entities  and relation in different embedding space. It means that two different spaces: entity space and relation space are modelled. Each entity is mapped into relation space. The translation construct is applied on the projected representation in the relationship space. The Figure-\ref{fig:my_label_3} presents the intuition behind TransR model.\\ 
To further refine the representation, a new model TransD was proposed by \cite{ji-etal-2015-knowledge}. TransR captures the possibility of relations and their embedding from the relation space. But the TransD, extends it to the entity space as well. Here Entity-relation pair is considered as the first-class object. 
\subsubsection{RotatE \cite{DBLP:journals/corr/abs-1902-10197}}
In knowledge graphs, often, the relation is symmetric/anti-symmetric, inversion and composition. e.g., "Marriage" is a symmetric relation, "My niece is my sister's daughter" is a composition, etc. The models discussed above are not capable of predicting these relations. The model proposed here is based on the intuition that the relation from head to tail is modeled as rotation in a complex plane. It is motivated by Euler's identity. 
    \begin{equation}
        e^{\iota\theta} = \sin \theta + \iota \cos \theta
    \end{equation}

\begin{figure}
    \centering
    \includegraphics[width=.9\linewidth]{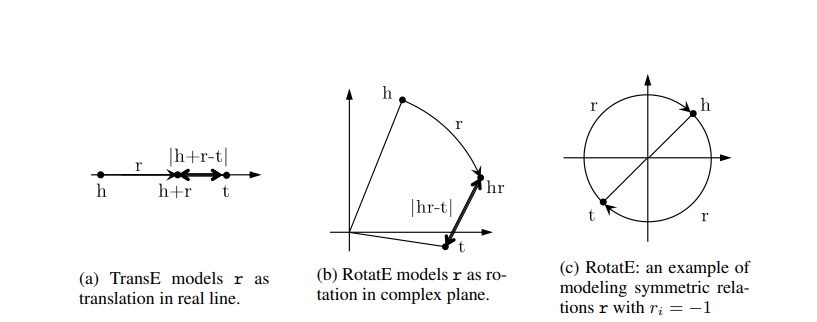}
    \caption{Representation of RotaE in comparison to TransE. In RotatE, the relationship between \textit{h} and \textit{t} is represented as angle of rotation. This image depicts the relationship in 1-dimension space. \citep{DBLP:journals/corr/abs-1902-10197} }
    \label{fig:my_label_4}
\end{figure}

\noindent
For a triplet \textit{(h,r,t)}, the relation among them can be represented as \textit{t = h $\circ$ r}. Where \textit{h, r, t} is the k-dimension embedding of the head, relation and tail, restricting the $|{r_i}|$ = 1. It means we are in the unit circle and $\circ$ represents element wise product. For each dimension $t_i = r_i h_i$ subjected to constraint $|{r_i}|$ = 1.
Under these condition the relationship is symmetric $\iff$  for all the values of i, $e^{0/\iota\theta} = \pm 1$, The relationship is inverse $\iff$ $r_j = \bar{r}_i$, i.e. both are the complex conjugate. Two relations are composite $\iff$ $r_j = r_i \circ r_k$. It means the relation $r_j$ can be obtained by a combined rotation of $r_i$ and $r_k$, $\theta_j = \theta_i + \theta_k$. The scoring function measures the angular distance.
\begin{equation}
    d_r(h,t) = \left\| h \circ r - t \right\|
\end{equation}
The Figure-\ref{fig:my_label_4} shows the comparison of RotaE and TransE. 
\subsubsection{HakE \cite{DBLP:journals/corr/abs-1911-09419}}
The approaches we have discussed so far fails to capture the semantic hierarchies. In HakE, the authors have proposed to model the hierarchy in the entities as concentric circles in polar coordinate. The entity with smaller radius belongs to higher up in the hierarchy. The angle between them represents the variation in the meaning. 
To represent a point on the circle, we should have $(r, \theta)$. Similarly, this model has two components, one to map the modulus and the other one to map the angle. The modulus part considers the depth of the tree as moduli. 
Let $h_m, t_m$ and $r_m$ are the representation in modulus space then $h_m \circ r_m = t_m$, where $h_m, t_m$ is a k-dim vector. 
The distance function is similar to RotatE with modification to consider only modulo part. 
\begin{equation}
    d_{r,m}(h_m,t_m) = \left\| h_m \circ r_m - t_m \right\|_2
\end{equation}
    
Similarly, the phase part can be formulated as 
$(h_p + r_p)mod 2\pi = t_p$ where $h_p, r_p, t_p \in [0,2\pi)^k$. The distance function is 
\begin{equation}
    d_{r,p}(h_p,t_p) = \left\|\sin ( (h_p \circ r_p - t_p)/2) \right\|_1
\end{equation}
By combining both the part, an entity can be mapped in the polar coordinate space.

\subsection{Semantic Matching Models}
Semantic Matching is one of the core tasks in Natural Language Processing. As we have seen that Translational Distance Models use distance-based scoring functions to calculate the similarity between the different entities and relations thus build the embedding accordingly. On the other hand, Semantic Matching Models use similarity-based scoring function. There are several Knowledge Graph Embedding algorithms comes under this model. Some of the algorithms are described below. 

\subsubsection{RESCAL \cite{nickel2011three}}
RESCAL follows Statistical Relational Learning Approach which is based on a Tensor Factorization model that takes the inherent structure of relational data into account. A Tensor\cite{kolda2009tensor} is a multidimensional array. More formally we can say that a first order Tensor is a vector, second order Tensor is a matrix and, Tensor with more than two order is called as higher order Tensor. Tensor Factorization is expressing a Tensor as a sequence of elementary operations acting on other, often a simpler Tensors.  Statistical Relational Learning inherits from Probability Theory and Statistics to address uncertainty and complexity of relational structures. \\
\cite{nickel2011three} models the Knowledge Graph triplet of the form (head, relation, tail) into three-way tensor, $\mathcal{X}$ as shown in  Figure 5.

\begin{figure}[htb]
    \includegraphics[width=.9\linewidth]{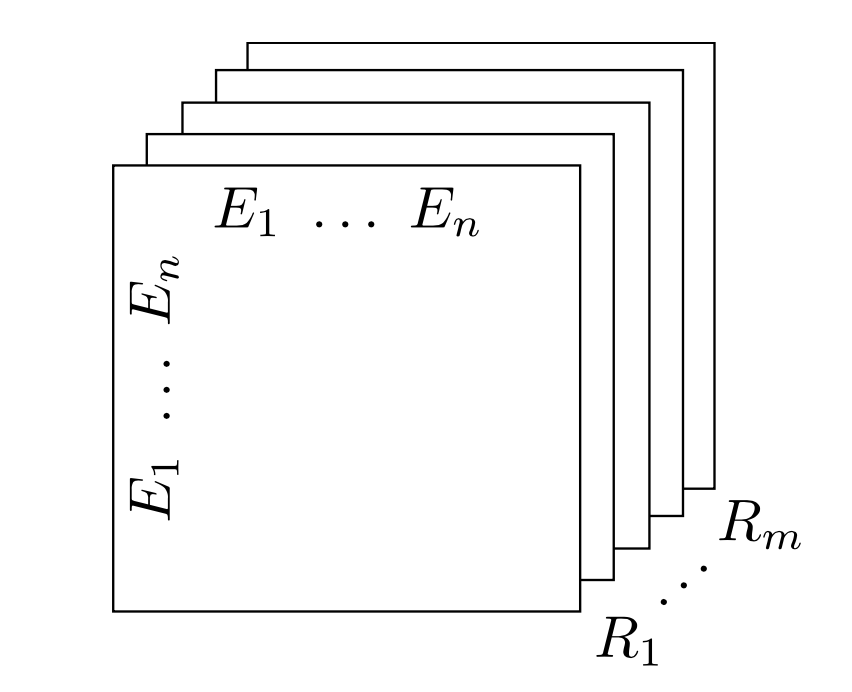}
    \caption{Tensor model for relational data. $E_{1} ... E_{n}$ denote the entities and $R_{1} ... R_{m}$ denote the relation in the domain\cite{nickel2011three} }
    \label{fig: Menu}
\end{figure}

\noindent
In  $\mathcal{X}$, two modes holds the concatenated entities (head and tail), and the third mode holds relation (relation). A Tensor entity $\mathcal{X}_{ijk}$ = 1  denotes that there exist a relation and if $\mathcal{X}_{ijk}$ = 0 denotes that there is unknown relation. It is assumed that data is given as n * n * m Tensor. Where n is the number of entities and m is the number of relations. RESCAL "explains triples via pairwise interaction of latent features". It performs the rank-r factorization on each slice of $\mathcal{X}$ (relational data) and the score of a fact (head, relation, tail ) is given by the following bi-linear function.
\[f_{r}(h,t) = \textbf{h}^T\textbf{M}_r\textbf{t} = \sum_{i=0}^{d-1}\sum_{j=0}^{d-1} [\textbf{M}_r]_{ij} .[\textbf{h}]_i . [\textbf{t}]_j\]
where h,t $\in$ $\mathbb{R}^d$ are vector representation of entities, and $M_{r}$ $\in$ $\mathbb{R}^{d*d}$ is a matrix representation of $r^{th}$ relation. Thus from this equation we are able to calculating the score of the triple using the weighted sum of all the pairwise interactions between the latent features of the entities $h$ and $t$ as shown in the Figure 6.

\begin{figure}[htb]
    \includegraphics[width=.9\linewidth]{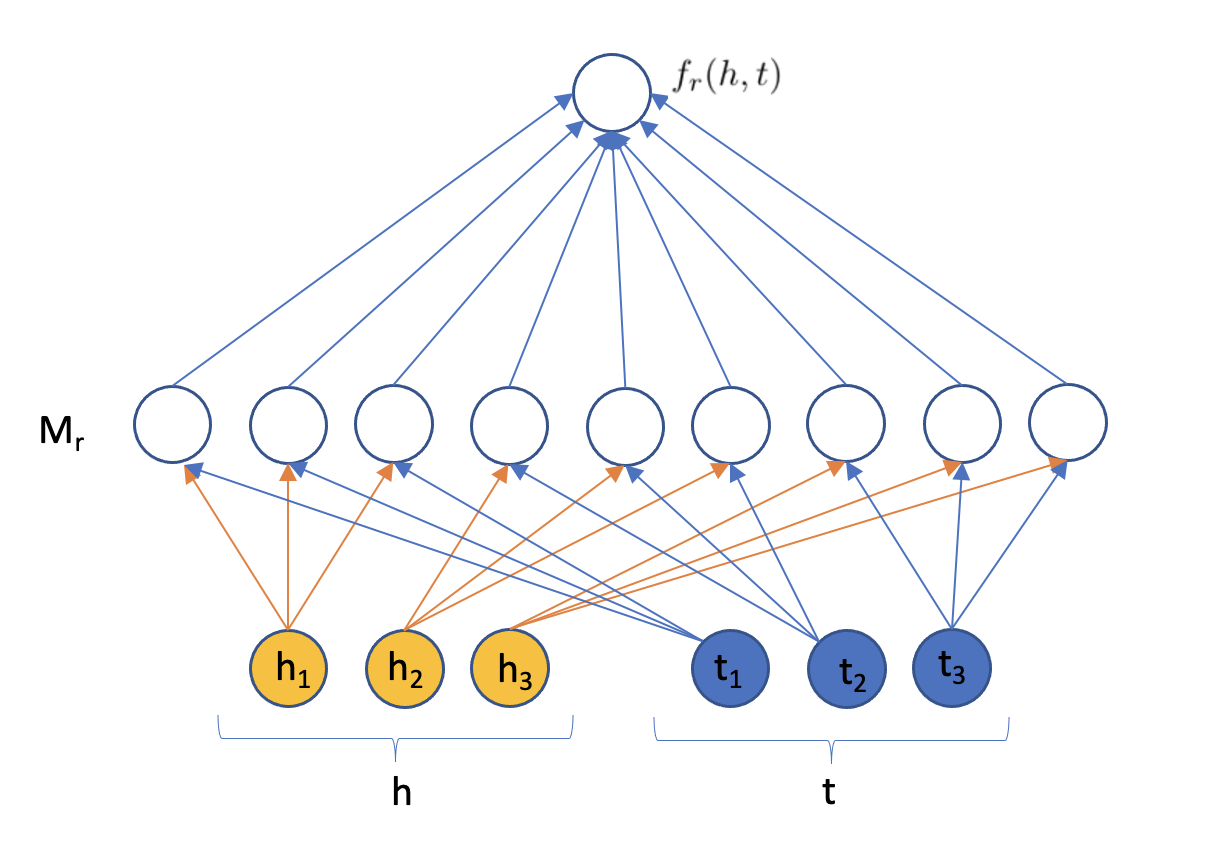}
    \caption{ RESCAL Representation. Here number of latent features for entities are 3 and number of latent features for relations are 3. \cite{Nickel_2016}  }
    \label{fig: Graph}
\end{figure}

\noindent
This method require $O(d^{2})$ parameters per relation and the space complexity of $O(nd + md^{2})$ where n is the number of entities and m is the number of relations.

\subsubsection{TATEC \cite{10.1007/978-3-662-44848-9_28}}
TATEC stands for \textit{Two And Three-way Embeddings Combination}. The main disadvantage of RESCAL is that it is a Three-way model which performs fairly good for relationships which occur frequently but it performs poor for the rare relationships and leads to major over-fitting. The issue of major over-fitting for rare relationships can be controlled by regularizing or reducing the expressivity of the model and former method is not feasible. The second method of reducing the expressivity is Two-way interaction which is implemented in TransE and SME. Two-way interaction approaches overperform the Three-way approaches on many datasets from which we can conclude that Two-way interactions are more efficient for the datasets and specially for those datasets which have more rare relationships. But the problem with the two-way interaction is that they are limited and are not able to represent all kind of relations with entities.\\
TATEC is a latent factor model which is capable of incorporating the high capacity Three-way model with well-controlled two-way interactions and take  the advantage of both of them. Since two-way and three-way models do not use the the same kind of data pattern and do not encode the same kind of information in the embedding. So, in TATEC during first stage they used two different embeddings and then combined and fine-tuned them in the later stage. The scoring function of TATEC is given by $f_{r}(h,t) = f_{r}^1(h,t) + f_{r}^2(h,t) $ which is a linear combination of bi-gram and tri-gram terms, where $f_{r}^1(h,t)$ is a two-way interaction score and $f_{r}^2(h,t)$ is a three-way interaction score. These can be calculates as follows:\\
1) Two-way interactions terms  can be given by:
\[  f_{r}^1(h,t) = \textbf{h}^T\textbf{r} + \textbf{t}^T\textbf{r} + \textbf{h}^T\textbf{D}\textbf{t} \]
\noindent
where D is the diagonal matrix shared across all the different relations and does not depend on input triple and r $\in$ $\mathbb{R}^d$ is a vector  that depends on relationships.\\
2) Three-way interactions terms  can be given by:
\[f_{r}^2(h,t) = \textbf{h}^T\textbf{M}_r\textbf{t} \]

\noindent
The final scoring function of TATEC is given by 
\[f_{r}(h,t) = \textbf{h}^T\textbf{M}_r\textbf{t} + \textbf{h}^T\textbf{r} + \textbf{t}^T\textbf{r} + \textbf{h}^T\textbf{D}\textbf{t}\]

\begin{figure}[htb]
    \includegraphics[width=.9\linewidth]{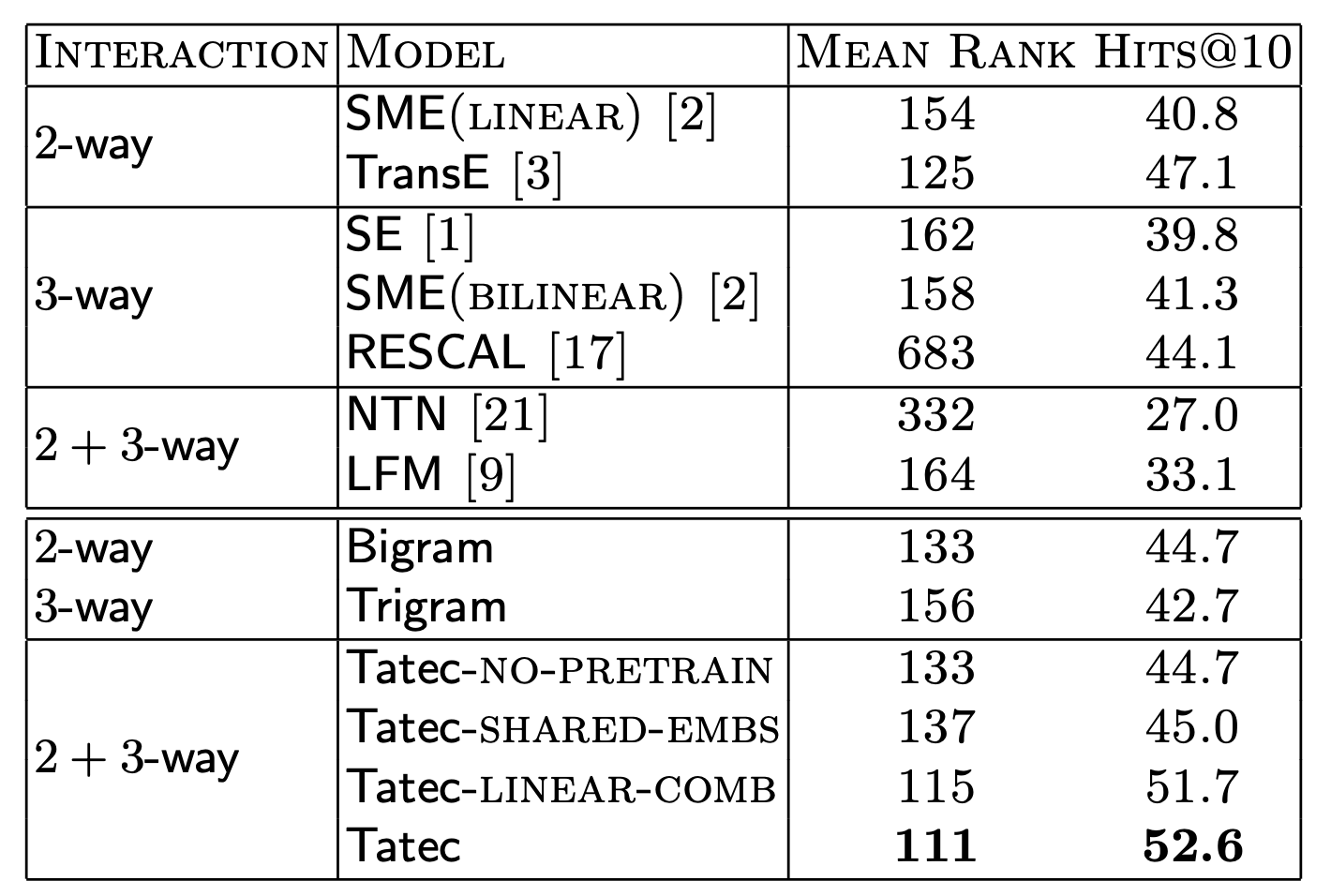}
    \caption{ Link prediction results \cite{10.1007/978-3-662-44848-9_28} }
    \label{fig: Graph}
\end{figure}

\noindent
Authors of \cite{10.1007/978-3-662-44848-9_28} compared this model to the other existing models such as RESCAL \cite{nickel2011three}, TransE \citep{Bordes}, LFM, SE and, SME for link prediction on FB15k dataset and as a result TATEC performs better than all other available models as shown in  Figure 7.

\noindent
Time complexity and the space complexity of TATEC is same as RESCAL as TATEC extends RESCAL. The time complexity of TATEC is $O(d^{2})$ parameters per relation and the space complexity of $O(nd + md^{2})$ where n is the number of entities and m is the number of relations.

\subsubsection{DistMult \cite{yang2014embedding}}
This model compares with the NTN neural model, TransE and  bi-linear models like RESCAL. The problem with NTN is that it is the most expensive model it incorporate both linear and bilinear relation operations. Similarly TransE parameterizes  the linear operations with one dimensional vectors. DistMult is a simplified RESCAL, which uses the basic bilinear scoring function.
\[f_{r}(h,t) = \textbf{h}^T\textbf{M}_r\textbf{t} \]

\noindent
These bilinear formulations are combined with different forms of regularization to make different models. In DistMult authors considered a simpler approach where they reduced the number of parameters by imposing restrictions on $M_{r}$ to be a diagonal matrix. This results in a simpler model and this model enjoys the same scalable properties of TransE as well as it achieves better performance over TransE. Thus the final scoring function is given as 
\[ f_{r}(h,t) = \textbf{h}^T diag(\textbf{r}) \textbf{t} = \sum_{i=0}^{d-1} [\textbf{r}]_{i} .[\textbf{h}]_i . [\textbf{t}]_i \]

\noindent
where for each relation r, r $\in \mathbb{R}^d$ is a vector that depends on relationships. 

\begin{figure}[htb]
    \includegraphics[width=.9\linewidth]{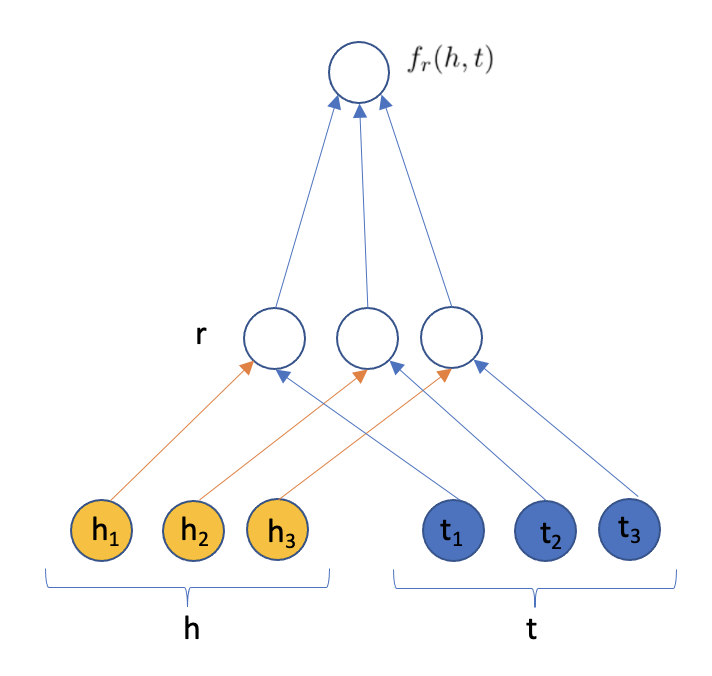}
    \caption{ DistMult simple Illustration }
    \label{fig: Graph}
\end{figure}

\noindent
Time complexity and the space complexity of DistMult is more efficient as compared to RESCAL or TATEC. The time complexity of DistMult is $O(d)$ parameters per relation and the space complexity of $O(nd + md)$ where n is the number of entities and m is the number of relations. Due to its over-simplified mature of the model, this model is not enough powerful for the use in case of general Knowledge Graphs because it is only able to work efficiently  with symmetric relations.

\subsubsection{HolE \cite{nickel2016holographic}}
HolE stands for Holographic Embedding. HolE tried to overcome the problem of Tensor Product used in RESCAL by using circular correlation. Tensor product uses pairwise multiplicative interactions between feature vectors which results in increase in dimensionality of the representation i.e., $\mathbb{R}^{d^2}$ thus increase the computational demand.
\[ \textbf{h} \circ \textbf{t} = \textbf{h} \otimes \textbf{t} \in \mathbb{R}^{d^2} \]
Where \textbf{a},\textbf{b} $\in \mathbb{R}^d$ are entity embeddings. Tensor products are very rich in capturing the interactions but are computational intensive. On the other hand HolE  use Circular Correlation which can be seen as compression of the Tensor Product. The main advantage of Circular Correlation over Tensor Product is that it won't increase the dimensionality of the representation.
\[ \textbf{h} \circ \textbf{t} = \textbf{h} * \textbf{t} \in \mathbb{R}^d \]
Where * : $ \mathbb{R}^d \times \mathbb{R}^d \rightarrow \mathbb{R}^d$ denotes the circular correlation. 
\[ [\textbf{h} * \textbf{t}]_{i} = \sum_{k=0}^{d-1} [\textbf{h}_{k}] . \textbf{t}_{(k+i)mod(d)}\]
The final score of the fact in  HolE is given by matching the compositional vector ($\textbf{h} * \textbf{t}$) with the relational representation,i.e.,

\[ f_{r}(h,t) = \textbf{r}^T  (\textbf{h} * \textbf{t} ) = \sum_{i=0}^{d-1} [\textbf{r}]_{i} \sum_{k=0}^{d-1}[\textbf{h}]_k . [\textbf{t}]_{(k+i)mod(d)} \]

\begin{figure}[htb]
    \includegraphics[width=.9\linewidth]{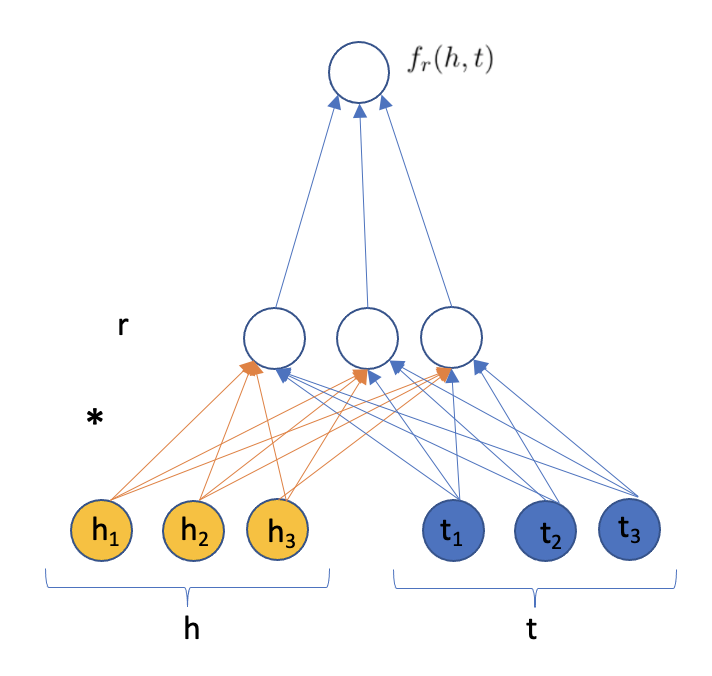}
    \caption{ HolE simple Illustration (It requires only d components) \cite{nickel2016holographic} }
    \label{fig: Graph}
\end{figure}

\noindent
So HolE is more efficient as compared to RESCAL or TransE. HolE take $O(d log d)$ parameters per relation and the space complexity of $O(nd + md)$ where n is the number of entities and m is the number of relations. Another advantage of HolE is that Circular Correlation is not commutative (  $\textbf{h} * \textbf{t} \neq \textbf{t} * \textbf{h}$ ) thus HolE is able to model asymmetric relations (directed graphs) with compositional representations which is not possible in RESCAL.

\subsubsection{ComplEx\cite{Trouillon2016}}
Knowledge graphs represent the relation between entities. The entities may be termed as subjects and objects respectively of a given relation. However not all relations may be present in a given KG. One of the application of KG is the ability to predict missing relations or entities. \\
Dot product of vector embedding of KG triplets is being successfully used for symmetric, reflexive, anti-reflexive and even transitive relations \citep{bouchard2015} however it can't be used for anti-symmetric relations. For example the relation capitalOf(New Delhi, India) is not symmetric since we cannot interchange subject and object entity in this relation therefore we need to have different embedding for an entity as subject and as object which increases the number of parameters.\\
Complex embedding facilitates joint learning of subject and object entities while preserving the asymmetry of the relation. It uses Hermitian dot product of embedding of subject entities and object entities. The Eigen Vector decomposition is used to identify a low rank diagonal matrix W such that there exists X = Re($EWE^{T}$) such that X has same sign pattern as Y. The low rank diagonal matrix W is then used to predict missing relations by applying Re($e_{s}^{T}W\bar{e}_{o}$).
\subsubsection{ANALOGY\cite{Liu2017}}
ANALOGY is based on a multiplicative model where the relation (s,r,o) is scored by multiplying the vector representations of of subject (s), relation (r) and object (o). The relation present in the knowledge graph are expected to have higher score i.e. $\phi(s,r,0)$ = $e_{s}^{T}W_{r}e_{o}$ will be high if the triplet (s,r,o) exists. \\
An example of analogy may be branch:tree::petal:leaf, in such an analogy, the relation $W_{r}$ "is part of" may be used to predict the missing entity from the analogy. The foundation of ANALOGY model is the linear maps of matrix representation of relations from the triplets present in the KG. Basically, the model uses the fact that there can be multiple paths to arrive at the entity from a given entity through a linear map of relations $W_{r1}, W_{r2}, W_{r3}$ and so on, and application of such relations in any sequence will give the same result. Hence such linear maps may be used to predict entities missing from the knowledge graph.\\
For example Let's consider the entities teacher (t), school (s), professor (p) and college (c). We have two relations in this set up: teachesAt (t,s) \& teachesAt(p,c) and juniorOf(t,p) \& juniorOf(s,c) therefore the relation between teacher and college is teachesAt*juniorOf = juniorOf*teachesAt. However such linear maps are feasible only if the commutative property holds for such relations. ANALOGY uses such linear maps to predict entities.

\subsection{Enrichment based embedding}
In the recent times, new emerging research areas are focusing on contextualized embedding. Under this, the entity under the consideration is enriched information from the neighbourhood information. A few notable approaches are Graph attention network (GAT) \citep{Velickovic2017a} based information enrichment. Two methods based on the KGAT \citep{Wei2019}, MMGAT \citep{Sun2020} has proposed models to embed contextual information for an entity. Under MMGAT, they proposed model to combine the embedding from multi-modal data with an attention framework, as adopted from GAT's attention framework. Both the frameworks, were using translation model to learn the representation after the enrichment. The new emerging research areas are try to learn the structural information as well as path based information, multi-modal data. There are other research areas in the embedding are: \textit{Text-enhanced embedding}, \textit{Logic-enhanced embedding}, \textit{Image-enhanced embedding} \citep{Bianchi2020KnowledgeGE} etc. 

\section{Applications of Knowledge graph embedding}

There are many applications of KG embedding learning methods. This section explores three of them namely- link prediction, triplet classification and recommender systems. 

\noindent
The first two are In-KG applications, which are conducted within the scope of the KG. The last is an example of Out-of-KG applications that scale to broader domains \citep{Wang2017}.

\subsection{Link Prediction}

The set of edges in a knowledge graph is a subset of \textit{Entities$\times$Relations$\times$Entities}. The link prediction task focuses on finding an entity that can be represented as a fact (edge) together with a given relation and entity i.e., \textit{(entity, relation, ?)} or \textit{(?, relation, entity)} where ? refers to the missing entity. For e.g. \textit{(New Delhi, isCapitalOf, ?)} or \textit{(?, isCapitalOf, India)}. Link prediction is a way of Knowledge graph augmentation \citep{paulheim2017knowledge}. It deduces missing information from the knowledge graph itself. 

\noindent
The datasets for LP are constructed by sampling from the original knowledge graph. Then, the links removed can be used in validation set or the test set \citep{bordes2013translating, dettmers2018convolutional}. The structure of such graphs play a vital role 
for improving the results, multiple source entities making learning effective and multiple destination entities making learning difficult \citep{Rossi_2021}.

\noindent
The LP models assigns a score to the triplet corresponding to each possible entity to fill the question mark (?). The triplets are then ranked by a function and entity corresponding to the lowest rank is predicted. If the predicted facts in the ranked predictions are already present in the Knowledge graph, they may or may not be excluded while calculating the ranks called raw and filtered rankings respectively \citep{bordes2013translating}. For e.g. if the training knowledge graph contains the fact that \textit{(Arjuna, isSonOf, Kunti)}, and the test query is \textit{(?, isSonOf, Kunti)}. The target answer is \textit{(Yudhishtra, isSonOf, Kunti)} and the system ranks  \textit{(Arjuna, isSonOf, Kunti)} and then \textit{(Yudhishtra, isSonOf, Kunti)}. The raw ranking of the triplet \textit{(Yudhishtra, isSonOf, Kunti)} will be two and filtered ranking will be one. 

\noindent
There are several tie breaking policies that are used by the ranking system. Assigning the \textit{minimum} or the \textit{
maximum} rank, or a \textit{random} or the \textit{average} rank to the targeted entity \citep{Rossi_2021}.

\noindent
The ranks obtained are used to compute metrics such as \textit{Mean Rank} (average of all the ranks), \textit{Mean Reciprocal Rank} (average of the inverse of ranks), or \textit{Hits@M} (proportion of ranks \textit{$\leq$ M}).

\subsection{Triple Classification}
Triple Classification is the problem of identifying whether a given triple is correct. It aims to give a yes or no answer to questions such as \textit{is New Delhi capital of India?} which can be written in the form of a triple \textit{(New Delhi, isCapitalOf, India)} \citep{socher2013reasoning}.

\noindent
A scoring function is used to calculate score of a triple similar to the link prediction. If the score is greater than a certain threshold, then it is considered a fact else a wrong triple \citep{Wang2017}. 

\noindent
Both the classical methods, such as \textit{micro} and \textit{macro averaging}, and ranking methods such as \textit{Mean Rank} are used as evaluation metrics \citep{guo2016jointly}. 

\subsection{Recommender Systems}
Recommender system (RS) assists the user in an environment where multiple options are available by providing a certain ordering of choices that the recommendation algorithm infers. This inference can be based on the similarity of the choices and behaviour pattern of different users. This type of recommendation methods falls into the domain of collaborative filtering methods \citep{adomavicius2005toward}. 

\noindent
The CF methods suffer from problems of Data sparsity and cold start. Data sparsity arises from the fact that only a small proportion of items are rated by the users and most options have only limited feedback from the users. Cold start problem is the problem of having no historical data about the new users and options. To deal with these problems different types of side information about a user and item are utilized by the RS \citep{Sun_2019}.

\noindent
KG is utilised for side information in CF. It acts as a heterogeneous graph that represent entities as nodes and relation as edges. The KG connects various entities via latent relationships and also provide explainability in recommendations \citep{Wang_2018}.

\noindent
The KG embedding based methods for RS use two modules - Graph embedding and Recommendation Module. The way that these modules are coupled lead to categorization of embedding based methods in a). two stage learning methods, b). joint learning method and c). multi task learning method\citep{Guo_2020}. 

\noindent
Two stage learning methods first uses graph embedding module to obtain the embeddings using various KG algorithms and then use recommendation module to infer. The advantages of this method lies in its simplicity and scalability but since the two modules are loosely coupled the embeddings might not be suitable for recommendation tasks.

\noindent
Joint learning methods train both the modules in an end to end fashion. Thus, recommendation module guides the training in graph embedding layer. 

\noindent
Multi task learning method train the recommendation module with the guidance of KG related task such as KG completion. The primary intuition behind this method is that the bipartite graph of user and item in recommendation task share structures with the corresponding KG entities.

\section{Summary}
Knowledge graphs provide an effective way of presenting real-world relationships. As a result Knowledge graphs have an inherent advantage w.r.t serving the information need. KG in itself is a growing area of research. KG embedding is a technique to represent all the components of the KG in vector form. These vectors represent the latent properties of the components of the graph. Various models for embedding methods are based on different combinations of vector algebra which present an interesting area of research. In this work, we have surveyed the embedding methods that started this active area of research, state-of-art models and the new frontiers which are being explored in the KG embedding. \\
The KG embedding methods progressed from translation-based models which are based on vector addition. In this work, we have presented how translation-based models improved over time to overcome shortcomings of the earlier models. While translation-based models used vector addition, semantic models can be clubbed together as multiplicative models. We have included the transition from basic semantic models to the more advanced semantic models which may be used to explain different types of real world relationships such as symmetric, anti-symmetric,inverse or composition.\\
New research areas have broaden the scope from structural embedding to more contextual embedding by encoding additional information in the learned representation. The latest area of research in this field is enrichment based embedding models. In this work, we have introduced those briefly. \\
Vector space representation has paved a way to use the information from Knowledge graph directly into the real world application. In this work, we have described a few real world applications of KG embedding such as link prediction, triple classification and recommender systems.

\bibliography{acl2020}
\bibliographystyle{acl_natbib}

\appendix

\end{document}